\documentclass[aps,showpacs,showkeys]{revtex4}%
\usepackage{amsfonts}
\usepackage{amsmath}
\usepackage{amssymb}
\usepackage{graphicx}%
\begin{document}
\title{Charge asymmetries in $e^{+}e^{-}\rightarrow\ \pi^{+}\ \pi^{-}\gamma$ at the
$\phi$ resonance.}
\author{A. Gallegos$^{(1)}$, J. L. Lucio$^{(2)}$, G. Moreno$^{(2)}$ and M.
Napsuciale$^{(2)}$ }
\affiliation{$^{(1)}$Departamento de Ciencias Exactas y Tecnolog\'{\i}a, Centro 
Universitario de los Lagos, Universidad de Guadalajara, Enrique D\'{\i}az de 
Le\'{o}n 1144, Colonia Paseos de la Monta\~{n}a, 47460, Lagos de Moreno, Jalisco, M\'{e}xico}

\affiliation{$^{(2)}$Departamento de F\'{\i}sica, Divisi\'{o}n de Ciencias e
Ingenier\'{\i}as, Universidad de Guanajuato, Campus Le\'{o}n, Lomas del Bosque
103, Fraccionamiento Lomas del Campestre, 37150, Le\'{o}n, Guanajuato, M\'{e}xico.}

\begin{abstract}
\noindent We consider the forward-backward pion charge asymmetry for the
$e^{+}e^{-}\rightarrow\pi^{+}\pi^{-}\gamma$ process. At tree level we consider 
bremsstrahlung and double resonance contributions. Although the latter 
contribution is formally sub-leading, it is enhanced at low dipion 
invariant mass due to $\rho$ resonant effects. We consider also four 
alternative models to describe the final state radiation at the loop 
level: Resonance Chiral Perturbation Theory, Unitarized Chiral 
Perturbation Theory, Kaon Loop Model and Linear Sigma Model. 
The last three models yield results
compatible with experimental data. The Kaon Loop Model requires an 
energy dependent phase to achieve the agreement.

\end{abstract}
\maketitle

\section{Introduction}

\noindent The nature of low mass scalar mesons nonet is a long-standing
puzzle. The $\phi$ radiative decays are expected to provide information about
the $f_{0}\left(  980\right)  $ and $a_{0}\left(  980\right)  $ scalar mesons.
Unfortunately data reported by the KLOE collaboration on the $\phi$ decays to
$f_{0}\gamma$ \cite{KLOEf0} and $a_{0}\gamma$ \cite{KLOEa0} -- with $\pi
^{0}\pi^{0}\gamma$ and $\pi^{0}\eta\gamma$ final states respectively--
together with results for the $\phi\rightarrow\pi^{+}\pi^{-}\gamma$ process,
including the $f_{0}\gamma$ as intermediate state \cite{KLOE}, are not
conclusive. In the latter work, results on the forward-backward asymmetry as a
function of the $\pi^{+}\pi^{-}$ invariant mass are presented. The asymmetry
is sensitive to the mechanisms involved in the final state radiation
\cite{kuhn01} and it provides information on the pion form factor \cite{PSV}.
Related work on the reaction $e^{+}e^{-}\rightarrow\ \pi^{+}\ \pi^{-}\gamma$
has been done aiming to elucidate the partonic structure of pions \cite{pire}.

\noindent The asymmetry requires a non vanishing interference between initial
(ISR) and final (FSR) state radiation, the latter being strongly model
dependent \cite{kuhn02}. The invariant amplitude for the $e^{+}e^{-}%
\rightarrow\pi^{+}\pi^{-}\gamma$ process can be parameterized in terms of
three independent Lorentz structures and thus the model dependence in FSR can
be included in three scalar functions $f_{i}$ \cite{Giulia}. The final state
radiation has been calculated in different models. The simplest approximation
has been named scalar QED \cite{kuhn02,kuhn03} and it actually includes the
$\rho$ contributions to the pion form factor. The contribution of scalars
($f_{0}(980)$ and $\sigma$) have been also considered using a point-like $\phi
f_{0}\gamma$ interaction, in the so called "no-structure" model
\cite{kuhn02,kuhn03}. Later on, the tree level bremsstrahlung of final pions
was calculated \cite{Giulia,PSV,Giulia2} within Resonance Chiral Perturbation
Theory ($R\chi PT$) \cite{EGPR}. In particular, in \cite{Giulia2} sub-leading
intermediate vector mesons contributions like $e^{+}e^{-}\rightarrow
\phi\rightarrow\rho^{\pm}\pi^{\mp}\rightarrow\pi^{+}\pi^{-}\gamma$, named
double resonance contributions, were incorporated.

\noindent The aim of this paper is to work out the one loop predictions for
$e^{+}e^{-}\rightarrow\pi^{+}\pi^{-}\gamma$ at the $\phi$ resonance using four
alternative models, namely $R\chi PT$, Unitarized Chiral Perturbation Theory
($U\chi PT$) \cite{OO} (containing actually a resumation of loops), Linear
Sigma Model $\left(  LSM\right)  $ \cite{Levy,Simon} and the so-called
"kaon-loop" model ($KLM$) \cite{LN}. In each case we add the tree level
contributions from bremsstrahlung of pions and the intermediate double
resonance, both proposed in \cite{Giulia2}. We report the forward-backward
pion charge asymmetry and compare our results with KLOE data.
 
\noindent The paper is organized as follows: Section II includes the general
formalism to describe the $e^{+}e^{-}\rightarrow\pi^{+}\pi^{-}\gamma$ process.
In section III we derive the scalar functions $f_{i}$ that characterize the
$R\chi PT$, $LSM$, $U\chi PT$ and $KLM$ contributions, including the tree
level bremsstrahlung and double resonance exchange. In section IV we present
the numerical results and compare them with data. Finally, conclusions are
given in section V.

\section{General formalism}

\noindent We are interested in the process
\begin{equation}
e^{-}\left(  p_{1}\right)  e^{+}\left(  p_{2}\right)  \rightarrow\pi
^{+}\left(  p_{+}\right)  \pi^{-}\left(  p_{-}\right)  \gamma\left(
k,\epsilon\right)  . \label{reaccion}%
\end{equation}
For completeness, in order to introduce our notation and conventions, in this
section we include the basic equations used to describe the process. To this
end, we follow the formalism developed in Ref.\cite{Giulia}. The invariant
amplitude $M$ includes the initial state radiation $M_{ISR}$, and final state
radiation $M_{FSR}$, \textit{i.e. } $M=M_{ISR}+M_{FSR}$, with
\begin{align}
M_{ISR}  &  =-\frac{e}{q^{2}}L^{\mu\nu}\epsilon_{\nu}^{\ast}l_{\mu}F_{\pi
}\left(  q^{2}\right)  ,\text{ \ \ }\\
\text{\ \ }M_{FSR}  &  =\frac{e^{2}}{s}J_{\mu}M_{F}^{\mu\nu}\epsilon_{\nu
}^{\ast}, \label{FSR}%
\end{align}
where $F_{\pi}\left(  q^{2}\right)  $ denotes the pion electromagnetic form
factor, $\epsilon_{\nu}$ is the photon polarization vector and the tensor
$M_{F}^{\mu\nu}$ describes the photon radiation from the final state. The
lepton currents are given by
\begin{align}
L^{\mu\nu}  &  =e^{2}\overline{u}_{s_{2}}\left(  -p_{2}\right)  \times\left[
\gamma^{\nu}\frac{\left(  -\not p  _{2}+\not k  +m_{e}\right)  }{t_{2}}%
\gamma^{\mu}+\gamma^{\mu}\frac{\left(  \not p  _{1}-\not k  +m_{e}\right)
}{t_{1}}\gamma^{\nu}\right]  \times u_{s_{1}}\left(  p_{1}\right)  ,\\
J_{\mu}  &  =e\overline{u}_{s_{2}}\left(  -p_{2}\right)  \gamma_{\mu}u_{s_{1}%
}\left(  p_{1}\right)  .
\end{align}

\noindent The electron and positron spinors are $u_{s_{1}}\left(
p_{1}\right)  $ and $\overline{u}_{s_{2}}\left(  -p_{2}\right)  $
respectively. In terms of the external particles' four-momenta, the following
variables are introduced $Q=p_{1}+p_{2},q=p_{+}+p_{-}$,$l=p_{+}-p_{-}$ and
five independent Lorentz scalars are defined
\begin{align}
s &  \equiv Q^{2}=2p_{1}\cdot p_{2},\nonumber\\
t_{1} &  \equiv\left(  p_{1}-k\right)  ^{2}=-2p_{1}\cdot k,\nonumber\\
t_{2} &  \equiv\left(  p_{2}-k\right)  ^{2}=-2p_{2}\cdot k,\\
u_{1} &  \equiv l\cdot p_{1},u_{2}\equiv l\cdot p_{2},\nonumber
\end{align}
where the electron mass has been neglected. The differential cross section is
\begin{equation}
d\sigma=\frac{1}{2s\left(  2\pi\right)  ^{5}}\int\delta^{4}\left(  p_{1}%
+p_{2}-p_{-}-p_{+}-k\right)  \times\frac{d^{3}p_{+}}{2E_{+}}\frac{d^{3}p_{-}%
}{2E_{-}}\frac{d^{3}k}{2\omega}\overline{\left\vert M\right\vert ^{2}%
},\label{cross section}%
\end{equation}
with $p_{+}=\left(  E_{+},\mathbf{p}_{+}\right)  ,p_{-}=\left(  E_{-}%
,\mathbf{p}_{-}\right)  $, $k=\left(  \omega=\left\vert k\right\vert
,\mathbf{k}\right)  $ and $\overline{\left\vert M\right\vert ^{2}}$ is the
squared invariant amplitude, averaged over initial lepton polarizations
\footnote{Ref. \cite{Giulia} uses $\overline{u}_{s^{\prime}}\left(  p\right)
u_{s}\left(  p\right)  =-\overline{u}_{s^{\prime}}\left(  -p\right)
u_{s}\left(  -p\right)  =2m_{e}\delta_{ss^{\prime}}$ and $\sum_{\text{polar.}%
}\epsilon_{\rho}^{\ast}\epsilon_{\sigma}=-g_{\rho\sigma}$.}. The most general
form of the FSR tensor $M_{F}^{\mu\nu}$ is \cite{Giulia}%
\begin{equation}
M_{F}^{\mu\nu}=f_{1}\tau_{1}^{\mu\nu}+f_{2}\tau_{2}^{\mu\nu}+f_{3}\tau
_{3}^{\mu\nu},\label{tensor}%
\end{equation}
where the $\tau_{i}^{\mu\nu}$ are three independent gauge invariant tensors
which are dictated by parity, charge conjugation, crossing symmetry and gauge
invariance
\begin{align}
\tau_{1}^{\mu\nu} &  =k^{\mu}Q^{\nu}-g^{\mu\nu}k\cdot Q,\nonumber\\
\tau_{2}^{\mu\nu} &  =k\cdot l\left(  l^{\mu}Q^{\nu}-g^{\mu\nu}k\cdot
l\right)  +l^{\nu}\left(  k^{\mu}k\cdot l-l^{\mu}k\cdot Q\right)  ,\\
\tau_{3}^{\mu\nu} &  =Q^{2}\left(  g^{\mu\nu}k\cdot l-k^{\mu}l^{\nu}\right)
+Q^{\mu}\left(  l^{\nu}k\cdot Q-Q^{\nu}k\cdot l\right)  .\nonumber
\end{align}
The scalar functions $f_{i}\equiv f_{i}\left(  Q^{2},k\cdot Q,k\cdot l\right)
$ are either even $\left(  f_{1,2}\right)  $ or odd $\left(  f_{3}\right)  $
under the change of sign of the argument $k\cdot l$. Our first task will be to
determine these scalar functions $f_{i}$ for $R\chi PT$, $U\chi PT$, $LSM$ and
$KLM$ in order to add it later to the tree level bremsstrahlung and double
resonance exchange \cite{Giulia2}.

\noindent The pair of pions produced in (\ref{reaccion}) differ in charge
conjugation, depending if the photon is emitted from the initial or from the
final state, while the former is odd under charge conjugation the latter is
even. So, any interference between the two amplitudes is odd under charge
conjugation and gives rise to a charge asymmetry. The forward-backward charge
asymmetry is defined as
\begin{equation}
A=\frac{N(\theta_{\pi^{+}}> 90^{\circ})-N(\theta_{\pi^{+}}< 90^{\circ})}
{N(\theta_{\pi^{+}}> 90^{\circ})+N(\theta_{\pi^{+}}< 90^{\circ})},
\label{asimetria2}%
\end{equation}
where $\theta_{\pi^{+}}$ is the $\pi^{+}$ polar angle, which is measured with
respect to the incident electron momentum. It should be clear that the
asymmetry depends strongly on the experimental conditions, in particular on
the cutoff polar angle and the minimal photon energy that can be measured.

\section{FSR Models}

\subsection{Bremsstrahlung}

\begin{figure}[t]
\centering \includegraphics[width=12cm,height=9.5cm]{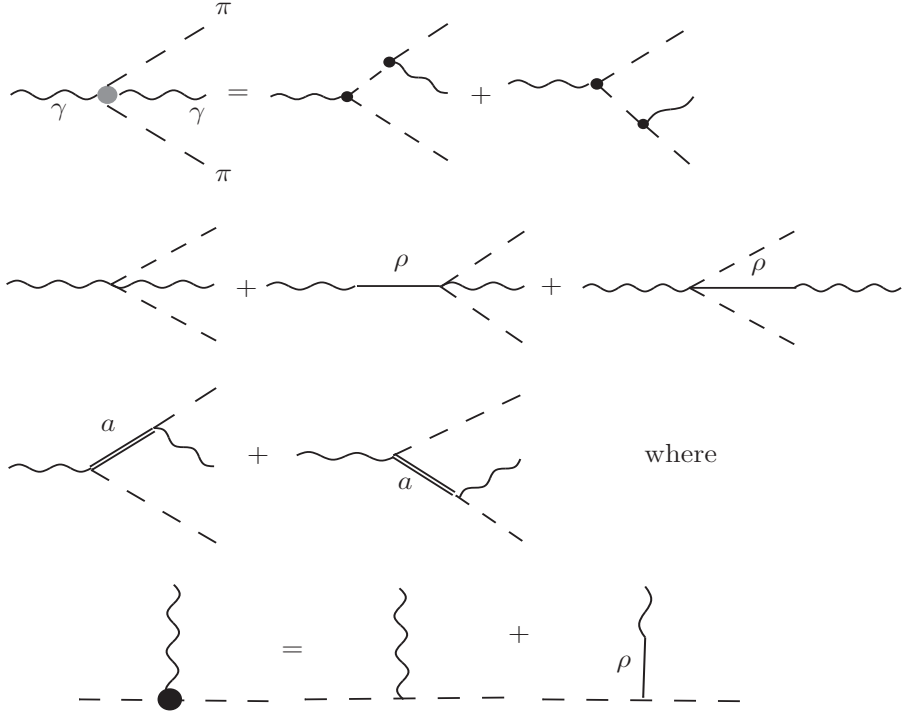}\caption{Feynman
diagrams for the bremsstrahlung, $a$ corresponds to $a_{1}(1260)$}%
\label{Bremms}%
\end{figure}

\noindent Before discussing the $\phi$ decay models at the loop level, we
first consider the bremsstrahlung of the final pions. The corresponding
Feynman diagrams are shown in Fig. (\ref{Bremms}) and the amplitude was
calculated in \cite{Giulia2}. The functions $f_{i}$ for this contribution are
given by equations (11) to (20) in \cite{Giulia2}.

\subsection{Double resonance contribution}

\begin{figure}[t]
\centering \includegraphics[width=12cm,height=2.5cm]{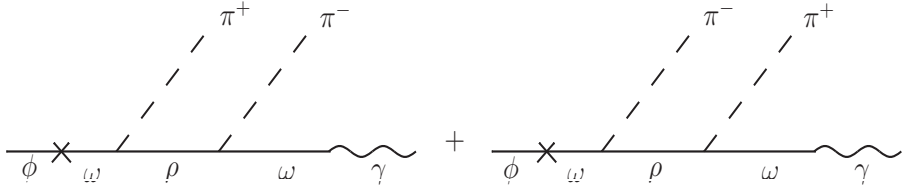}\caption{Feynman
diagrams for the double resonance contribution}%
\label{double}%
\end{figure}

\noindent The double resonance contribution $e^{+}e^{-}\rightarrow
\phi\rightarrow\rho^{\pm}\pi^{\mp}\rightarrow\pi^{+}\pi^{-}\gamma$ is
described by the diagrams shown in Fig. (\ref{double}). This process was
calculated in \cite{Isidori} and used in \cite{Giulia2} (equations (26) to
(28)), explicitly the $f_{i}$ functions are\footnote{We have included a 2
factor in $f_{3}^{VMD}$. The $f_{i}$ functions in \cite{Giulia2} are deduced
using the functions $L_{\mu\nu}^{\left(  i\right)  }$ defined in
\cite{Isidori} which are different from the $\tau_{i}^{\mu\nu}$ used here.} :%
\begin{align}
f_{1}^{VMD} &  =-\frac{1}{4\pi\alpha s}\left(  \left(  -1+\frac{3}{2}%
x+\frac{4m_{\pi}^{2}}{s}\right)  \left(  g\left(  x_{1}\right)  +g\left(
x_{2}\right)  \right)  +\frac{1}{4}\left(  x_{1}-x_{2}\right)  \left(
g\left(  x_{1}\right)  -g\left(  x_{2}\right)  \right)  \right)  ,\nonumber\\
f_{2}^{VMD} &  =-\frac{1}{4\pi\alpha s^{2}}\left(  g\left(  x_{1}\right)
+g\left(  x_{2}\right)  \right)  ,\label{factors VME isidori}\\
f_{3}^{VMD} &  =-\frac{1}{4\pi\alpha s^{2}}\left(  g\left(  x_{1}\right)
-g\left(  x_{2}\right)  \right)  ,\nonumber
\end{align}
where%
\begin{align}
g\left(  x\right)   &  =\frac{eg_{\rho\pi}^{\phi}g_{\pi\gamma}^{\rho}%
}{4F_{\phi}}\frac{m_{\phi}^{2}e^{i\beta_{\rho}}e^{i\beta_{\omega\phi}}%
}{s-m_{\phi}^{2}+im_{\phi}\Gamma_{\phi}}\frac{s^{2}\Pi_{\rho}^{VMD}}{\left(
1-x\right)  s-m_{\rho}^{2}+im_{\rho}\Gamma_{\rho}\left(  \left(  1-x\right)
s\right)  },\label{gx}\\
x_{1,2} &  =\frac{2p_{+,-}\cdot Q}{s},\qquad x=2-x_{1}-x_{2,}\nonumber
\end{align}
with the following values for the involved parameters%
\begin{align*}
g_{\rho\pi}^{\phi} &  =0.811\text{ GeV}^{-1},\qquad g_{\pi\gamma}^{\rho
}=0.295\text{ GeV}^{-1},\\
F_{\phi} &  =42.5,\qquad\Pi_{\rho}^{VMD}=0.58195,\\
\beta_{\rho} &  =32.996^{\circ},\qquad\beta_{\omega\phi}=163^{\circ}.
\end{align*}
As we shall see below, this contribution is very important in the description
of the charge asymmetries at low dipion invariant mass. It has also been
calculated recently \cite{Roca:2009zy} using the Lagrangian
\begin{equation}
\mathcal{L}=\frac{G}{\sqrt{2}}\epsilon_{\mu\nu\alpha\beta}tr\left(
\partial^{\mu}V^{\nu}\partial^{\alpha}V^{\prime\beta}\Phi\right)
-4f^{2}egA^{\mu}tr\left(  QV_{\mu}\right)  +\Theta\phi_{\mu}\omega^{\mu
}\label{VME}%
\end{equation}
where $g=-4.41$, $G=\frac{3g^{2}}{4\pi^{2}f}=0.016$ MeV$^{-1},$
$Q=diag\left\{  2/3,-1/3,-1/3\right\}  $ and $e$ is taken positive. The
$\phi-\omega$ mixing strength is given by $\widetilde{\varepsilon}%
=\frac{\Theta}{M_{\phi}^{2}-M_{\omega}^{2}}=0.059\pm0.004.$\ In this scheme we
find the $f_{i}$ scalar functions as
\begin{align}
f_{1} &  =\alpha\left[  \left(  l^{2}+k\cdot Q+2k\cdot l\right)  D_{\rho
}\left(  P\right)  +\left(  l^{2}+k\cdot Q-2k\cdot l\right)  D_{\rho}\left(
P^{\prime}\right)  \right]  \nonumber\\
f_{2} &  =-\alpha\left[  D_{\rho}\left(  P\right)  +D_{\rho}\left(  P^{\prime
}\right)  \right]  \label{factors VME}\\
f_{3} &  =-\alpha\left[  D_{\rho}\left(  P\right)  -D_{\rho}\left(  P^{\prime
}\right)  \right]  \nonumber
\end{align}
where $P=\frac{1}{2}\left(  Q+k-l\right)  $, $P^{\prime}=\frac{1}{2}\left(
Q+k+l\right)  $ and $\alpha$ stands for
\[
\alpha=-\frac{4f^{4}g^{2}G^{2}\widetilde{\varepsilon}}{9\sqrt{2}M_{\omega}%
^{2}}D_{\phi}\left(  Q\right)  ,
\]
which coincides with results in \cite{Roca:2009zy} whenever $M_{V}=2|g|f.$
This relation is well satisfied numerically and it is possible to show that
these functions coincide with (\ref{factors VME isidori}) up to the phases
included in (\ref{gx}) which have a small effect on the asymmetry.

\subsection{Contributions from $R\chi PT$ and $U\chi PT$}

\begin{figure}[t]
\centering \includegraphics[width=12cm,height=9.5cm]{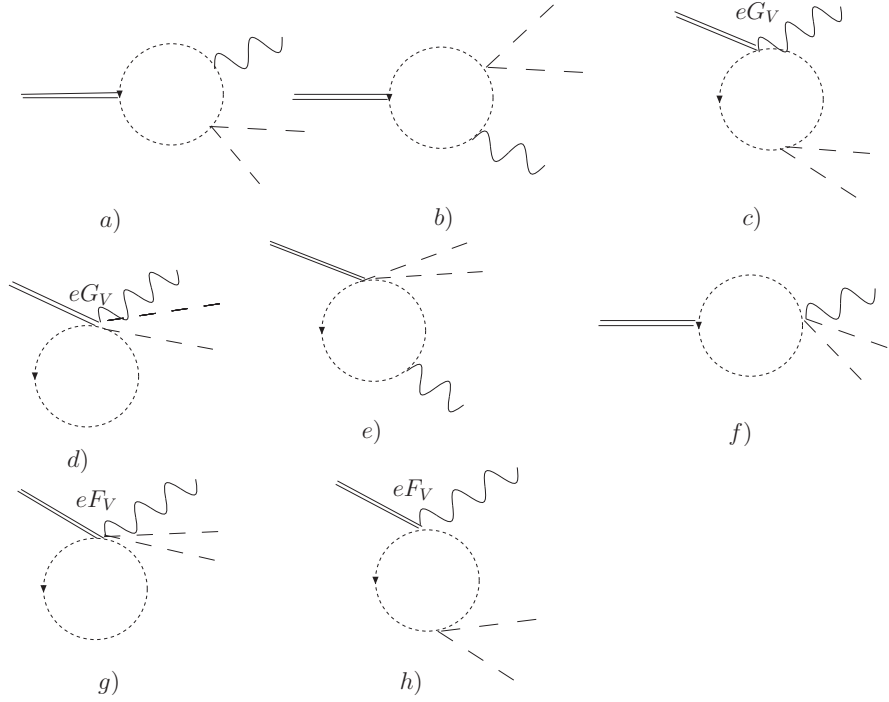}\caption{Feynman
diagrams for the $\gamma\phi\pi\pi$ vertex function in $R\chi PT$}%
\label{FD}%
\end{figure}

\noindent The amplitude for $e^{+}e^{-}\rightarrow\pi^{+}\pi^{-}\gamma$ at the
$\phi$ peak involves the $\gamma\phi\pi^{+}\pi^{-}$ vertex function with all
particles on-shell. This vertex function was calculated in the context of
$R\chi PT$ and $U\chi PT$ within the analysis of $e^{+}e^{-}\rightarrow\phi
\pi\pi$ for an \textit{off-shell} photon \cite{NSOV}. The relevant diagrams in
$R\chi PT$ \cite{EGPR} are shown in Fig. (\ref{FD}). These diagrams include
kaons in the loops, thus they involve the off-shell $K\overline{K}-\pi\pi$
amplitude. It was shown in \cite{NSOV} that, to leading order in the chiral
expansion, the contribution of diagrams $d,e,f,g$ cancels the off-shell
contributions of the $K\overline{K}-\pi\pi$ amplitude, entirely contained in
diagrams $a,b,c,h$, so that the calculation reduces to evaluate diagrams
$a,b,c,h$ with the $K\overline{K}-\pi\pi$ amplitude on- shell. This procedure
yields the $R\chi PT$ result for the $\gamma\phi\pi^{+}\pi^{-}$ vertex
function and we would expect it to reproduce experimental results at low
dipion invariant mass. However, due to the appearance of the widely discussed
light scalar resonance (the $\sigma$ meson), this expansion breaks down in
the scalar channel even at the dipion threshold.

\noindent The scalar poles can be generated unitarizing the leading order
meson-meson scattering amplitudes for definite isospin. Following \cite{OO},
the unitarized $K^{+}K^{-}-\pi^{+}\pi^{-}$ scattering amplitude is calculated
projecting onto the zero spin and isospin channel the leading order
meson-meson amplitude and performing a coupled channel analysis involving
iterations of all intermediate states in the $s$ channel.

\noindent As far as the calculation of the $e^{+}e^{-}\rightarrow\pi^{+}%
\pi^{-}\gamma$ amplitude is concerned, the scalar poles are incorporated
replacing the leading order on-shell $K^{+}K^{-}-\pi^{+}\pi^{-}$ amplitude by
the $K^{+}K^{-}-\pi^{+}\pi^{-}$ unitarized amplitude. For details of the
calculation we refer the interested reader to \cite{NSOV}, here we just quote
the result for $\phi\rightarrow\pi^{+}\pi^{-}\gamma$. The resulting amplitude
for $\phi(Q,\eta^{\alpha\nu})\rightarrow\pi^{+}\left(  p_{+}\right)  \pi
^{-}\left(  p_{-}\right)  \gamma\left(  k,\epsilon^{\mu}\right)  $ in $U\chi
PT$ is

\begin{align}
-i\mathcal{M}  &  =\frac{2}{\sqrt{3}}\frac{e}{2\sqrt{2}\pi^{2}m_{K}^{2}f^{2}%
}\frac{t_{K\pi}^{0}}{\sqrt{3}}\left[  G_{V}\left(  \widetilde{I}_{P}%
^{ab}(Q\cdot k\ g_{\mu\nu}-Q_{\mu}k_{\nu})\right)  Q_{\alpha}\right.
\nonumber\\
&  -\left.  \left(  G_{V}-\frac{F_{V}}{2}\right)  \frac{m_{K}^{2}}{4}%
g_{K}(q^{2})g_{\mu\nu}k_{\alpha}\right]  \eta^{\alpha\nu}\epsilon^{\mu},
\label{MUCHPT}%
\end{align}
where $q^{2}=(p_{+}+p_{-})^{2}$ and $t_{K\pi}^{0}$ denotes the unitarized
isoscalar scalar $K\overline{K}-\pi\pi$ amplitude. The factor $2/\sqrt{3}$ is
required to single out the $\pi^{+}\pi^{-}$ contribution in the isoscalar
$\pi\pi$ channel. The function $g_{K}(q^{2})$ is given by
\begin{equation}
g_{K}(q^{2})=-1+\log\frac{m_{K}^{2}}{\mu^{2}}+\sigma(q^{2})\log\frac
{\sigma(q^{2})+1}{\sigma(q^{2})-1}%
\end{equation}
with $\sigma(q^{2})=\sqrt{1-\frac{4m_{K}^{2}}{q^{2}}}$. Note that the
particular form of this function involves a regularization scheme as well as a
subtraction point. We use dimensional regularization and the value
$\mu=1.2\,GeV$ which reproduces the $f_{0}$ peak at $980MeV$ in the squared
meson-meson amplitudes in the scalar channel \cite{OO}. The loop integral is
given by
\begin{align}
\widetilde{I}_{P}^{ab}  &  =\frac{1}{2\left(  a-b\right)  }-\frac{2}{\left(
a-b\right)  ^{2}}\left[  f\left(  \frac{1}{b}\right)  -f\left(  \frac{1}%
{a}\right)  \right]  +\frac{a}{\left(  a-b\right)  ^{2}}\left[  g\left(
\frac{1}{b}\right)  -g\left(  \frac{1}{a}\right)  \right]  ,\label{loop}\\
f\left(  z\right)   &  =\left\{
\begin{array}
[c]{c}%
-\left[  \arcsin\left(  \frac{1}{2\sqrt{z}}\right)  \right]  ^{2}\text{
\ }z>\frac{1}{4}\\
\frac{1}{4}\left[  \ln\left(  \frac{n_{+}}{n_{-}}\right)  -i\pi\right]
^{2}\text{ \ \ \ }z<\frac{1}{4}%
\end{array}
\right.  ,\text{ \ }g\left(  z\right)  =\left\{
\begin{array}
[c]{c}%
\sqrt{4z-1}\arcsin\left(  \frac{1}{2\sqrt{z}}\right)  \text{ \ \ \ }z>\frac
{1}{4}\\
\frac{1}{2}\sqrt{1-4z}\left(  \ln\left\vert \frac{n_{+}}{n_{-}}\right\vert
-i\pi\right)  \text{ \ \ \ }z<\frac{1}{4}%
\end{array}
\right.  ,\nonumber\\
\text{ \ \ \ }a  &  =\frac{Q^{2}}{m_{K}^{2}},\text{ }b=\frac{q^{2}}{m_{K}^{2}%
},\text{ \ }n_{\pm}=\frac{1}{2}\left[  1\pm\sqrt{1-4z}\right]  .\nonumber
\end{align}
Results for $R\chi PT$ are obtained from Eq. (\ref{MUCHPT}) by replacing
$t_{K\pi}^{0}$ by the leading order on-shell interaction $V_{K\pi}=-\sqrt
{3}q^{2}/4f^{2}$. Using the propagator for a vector meson in the tensor
formalism we obtain the amplitude for $\gamma^{\ast}(Q,\mu)\rightarrow\pi
\pi\gamma(k,\nu)$ via the exchange of the vector meson $\phi$ as
\begin{equation}
-i\mathcal{M}=-\frac{ie^{2}F_{V}\sqrt{2}}{3}\frac{1}{Q^{2}-M_{\phi}%
^{2}+i\Gamma_{\phi}M_{\phi}}\left[  \left(  Q^{2}I+J\right)  \left(  k_{\mu
}Q_{\nu}-Q\cdot kg_{\mu\nu}\right)  \right]  \varepsilon^{\nu},
\end{equation}
where%
\begin{align}
I  &  =-\frac{G_{V}}{\sqrt{6}\pi^{2}m_{K}^{2}f^{2}}\frac{t_{K\pi}^{0}}%
{\sqrt{3}}\widetilde{I}_{P}^{ab},\qquad\\
J  &  =\frac{1}{\sqrt{6}\pi^{2}m_{K}^{2}f^{2}}\frac{t_{K\pi}^{0}}{\sqrt{3}%
}\left(  G_{V}-\frac{F_{V}}{2}\right)  \frac{m_{K}^{2}}{4}g_{K}(q^{2}).
\end{align}

\noindent In terms of this vertex function we can identify the final state
radiation invariant tensor $M_{F}^{\mu\nu}$ (see Eq. \ref{tensor})
\begin{equation}
M^{\mu\nu}=-ie^{2}(f_{1}\tau_{1}^{\mu\nu}+f_{2}\tau_{2}^{\mu\nu}+f_{3}\tau
_{3}^{\mu\nu})=-ie^{2}M_{F}^{\mu\nu}%
\end{equation}
with%
\begin{align}
f_{1}  &  =-\frac{1}{\sqrt{3}}\frac{F_{V}}{3f^{2}}\frac{1}{Q^{2}-M_{\phi}%
^{2}+i\Gamma_{\phi}M_{\phi}}\frac{t_{K\pi}^{0}}{\pi^{2}\sqrt{3}}\left(
\frac{Q^{2}}{m_{K}^{2}}G_{V}\widetilde{I}_{P}^{ab}-\frac{1}{4}\left(
G_{V}-\frac{F_{V}}{2}\right)  g_{K}(q^{2})\right)  ,\label{rescalar}\\
f_{2}  &  =0,\\
f_{3}  &  =0.
\end{align}
Notice that Eq.(\ref{rescalar}) contains a term with the combination
$G_{V}-\frac{F_{V}}{2}$ . This combination is small and it vanishes in the
context of Vector Meson Dominance \cite{GVFV}.

\subsection{The phenomenological Kaon Loop Model}

\begin{figure}[t]
\centering \includegraphics[width=12cm,height=2.5cm]{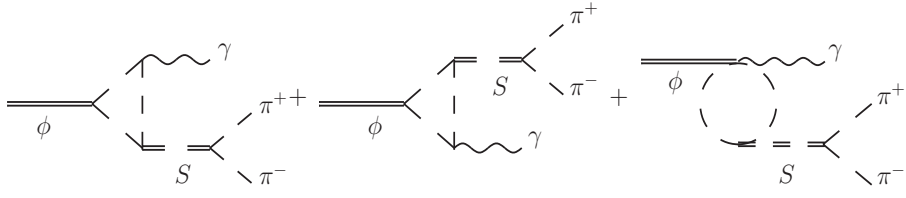}\caption{Feynman
diagrams for the kaon loop model}%
\label{loop01}%
\end{figure}

\noindent In this model the process under consideration proceeds through the
chain
\begin{equation}
e^{-}\left(  p_{1}\right)  e^{+}\left(  p_{2}\right)  \rightarrow
\phi\rightarrow S\left(  q\right)  \gamma\left(  k,\epsilon\right)
\rightarrow\pi^{+}\left(  p_{+}\right)  \pi^{-}\left(  p_{-}\right)
\gamma\left(  k,\epsilon\right)  ,
\end{equation}
with $S=f_{0},\sigma$. The corresponding Feynman diagrams of the $\phi$ decay 
are shown in figure (\ref{loop01}). The amplitude is
\begin{equation}
M_{\phi}=\frac{-ie^{2}}{s}A\,e\overline{\,v}\left(  p_{2}\right)  \gamma_{\mu
}u\left(  p_{1}\right)  \left(  Q^{\nu}k^{\mu}-Q\cdot kg^{\mu\nu}\right)
\epsilon_{\nu},\label{amplitud phi reescrita}%
\end{equation}
\noindent where we have defined
\begin{equation}
A=\frac{g_{s}g_{\phi}}{f_{\phi}}\frac{g_{f}}{2\pi^{2}m_{K^{+}}^{2}}%
\widetilde{I}_{P}^{ab}F_{\phi}\left(  s\right) \left[  
\sum_{S=f,\sigma}\frac{g_{S\pi^{+}\pi^{-}}g_{SK^{+}K^{-}}}
{D_{S}\left(  q^{2}\right)  }\right],
\end{equation}
with
\begin{equation}
F_{\phi}\left(  s\right)  =\frac{m_{\phi}^{2}}{s-m_{\phi}^{2}+i\sqrt{s}%
\Gamma_{\phi}},\hspace{1cm}D_{S}\left(  q^{2}\right)  =q^{2}%
-m_{S}^{2}+im_{S}\Gamma_{S},
\end{equation}
\noindent and $g_{\phi},f_{\phi}$ stand for the $\phi K^{+}K^{-}$ and 
$\phi\gamma$ couplings respectively (for details concerning the precise 
definition of these quantities we refer the reader to the appendix of 
Ref.\cite{LN}).  The kaon loop function $\widetilde{I}_{P}^{ab}$ is 
given in (\ref{loop}). It must be mentioned that the $KLM$ does not 
account for elastic (i.e. non-resonant) $K^{+}K^{-}\rightarrow\pi^{+}\pi^{-}$ 
scattering of kaons in the loops and the final pions. It has been shown in 
\cite{kuhn02} that this contribution is important in the interference 
between bremsstrahlung plus double resonance and the $KLM$. This elastic 
contribution was considered by the introduction of an energy dependent 
phase in the $KLM$ amplitude. Finally, the scalar functions for this model 
are obtained by comparing (\ref{amplitud phi reescrita}) with (\ref{FSR}) 
and (\ref{tensor}), in this way we  get%
\begin{align}
f_{1} &  =A,\nonumber\\
f_{2} &  =0,\label{factores}\\
f_{3} &  =0.\nonumber
\end{align}

\subsection{The Linear Sigma Model}

\begin{figure}[t]
\centering \includegraphics[width=12cm,height=6.0cm]{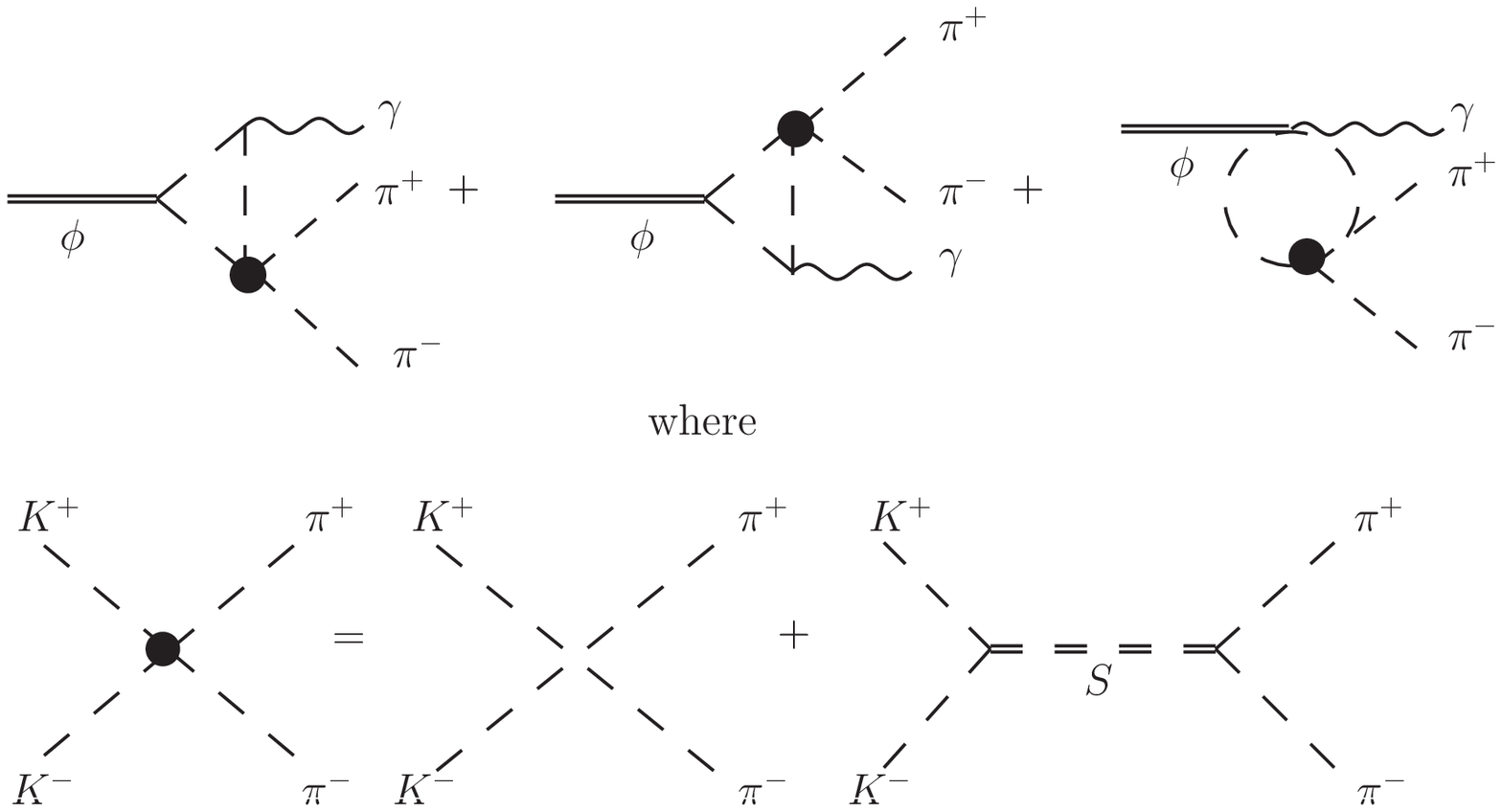}\caption{Feynman
diagrams for the linear sigma model}%
\label{loop02}%
\end{figure}

\noindent The calculation in this approach is similar to the $KLM$, the
difference arising from the treatment of the scalars. The Feynman diagrams are
shown in figure (\ref{loop02}). For the neutral pion case, the amplitude has
already been derived in \cite{Bramon} using the improved chiral loop approach.
Thus we can obtain the amplitude we are interested in just by making the
following replacement in the $KLM$ amplitude
\begin{equation}
\sum_{S=f,\sigma}\frac{g_{S\pi^{+}\pi^{-}}g_{SK^{+}K^{-}}}%
{D_{S}\left(  q^{2}\right)  }  \rightarrow\mathcal{A}\left(  K^{+}%
K^{-}\rightarrow\pi^{+}\pi^{-}\right)  _{L\sigma M}=\sqrt{2}\mathcal{A}\left(
K^{+}K^{-}\rightarrow\pi^{0}\pi^{0}\right)  _{L\sigma M}\text{,}
\label{replace}%
\end{equation}
\noindent where the amplitude for the meson scattering is given by
\begin{align}
\mathcal{A}\left(  K^{+}K^{-}\rightarrow\pi^{0}\pi^{0}\right)  _{L\sigma M}
&  =\frac{m_{\pi}^{2}-q^{2}/2}{2f_{\pi}f_{K}}+\frac{q^{2}-m_{\pi}^{2}}%
{2f_{\pi}f_{K}}\left[  \frac{m_{K}^{2}-m_{\sigma}^{2}}{D_{\sigma}\left(
q^{2}\right)  }\text{c}_{\phi_{S}}\left(  \text{c}_{\phi_{S}}-\sqrt{2}%
\text{s}_{\phi_{S}}\right)  \right. \nonumber\\
&  +\left.  \frac{m_{K}^{2}-m_{f_{0}}^{2}}{D_{f_{0}}\left(  q^{2}\right)
}\text{s}_{\phi_{S}}\left(  \text{s}_{\phi_{S}}+\sqrt{2}\text{c}_{\phi_{S}%
}\right)  \right]  ,
\end{align}
with $f_{K}=1.22f_{\pi}$, $\left(  \text{c}_{\phi_{S}},\text{s}_{\phi_{S}}\right)
\equiv\left(  \cos\phi_{S},\sin\phi_{S}\right)  $ and $\phi_{S}$ is the scalar
mixing angle in the strange-non-strange basis \cite{Simon}. The scalar
functions for this model can be obtained from (\ref{factores}) by making the
replacement in Eq. (\ref{replace}).

\section{Numerical results}

\noindent Numerical results are obtained using a Monte Carlo code where the
experimental conditions of the KLOE collaboration are included. Thus, for the
$\pi^{+}$ polar angle - defined respect to the electron beam - we considered
the range $45^{\circ}<\theta_{\pi^{+}}<135^{\circ}$. As far as the photon is
concerned we take $45^{\circ}<\theta_{\gamma}<135^{\circ}$ and assume
$E_{\gamma}>10$ MeV \cite{KLOE}.

\noindent Calculations in $U\chi PT$ and $R\chi PT$ involve parameters that
have already been fixed from meson phenomenology. We use the following values:
$G_{V}=53$ MeV, $F_{V}=154$ MeV, $f_{\pi}=93$ MeV and $\mu=1.2$ GeV \cite{OO}.
Concerning the $KLM$, a summary of the involved parameters is given in Table \ref{tabla}.
\bigskip

\begin{figure}[ptb]
\begin{center}
\includegraphics[height=3.3in,width=4.7in
]{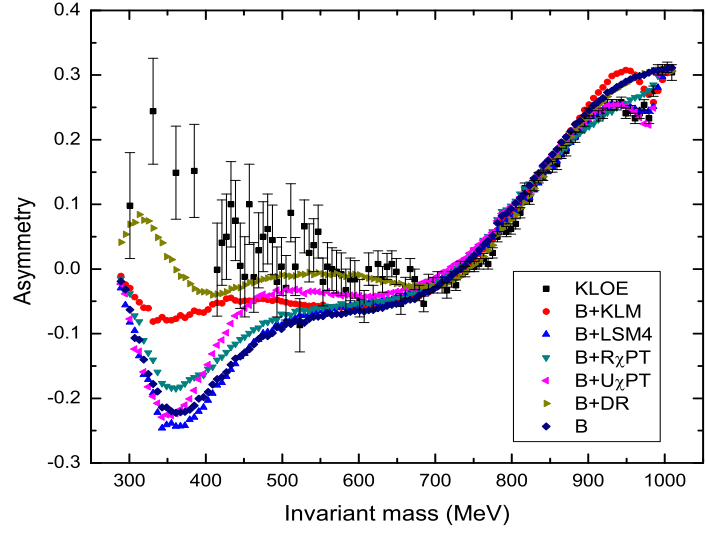}
\end{center}
\caption{Results for the forward backward asymmetry as predicted by the bremsstrahlung ($B$), 
bremsstrahlung plus double resonance ($B+DR$) and bremsstrahlung plus loop 
models for the kaon loop model ($B+KLM$), linear sigma model ($B+LSM4$), resonance chiral 
perturbation theory ($B+R\chi PT$) and unitarized chiral perturbation theory 
($B+U\chi PT$). Data is taken from Refs. \cite{KLOE,tesis}.}
\label{UCHPT}%
\end{figure}

\begin{figure}[ptb]
\begin{center}
\includegraphics[height=3.3in,width=4.7in
]{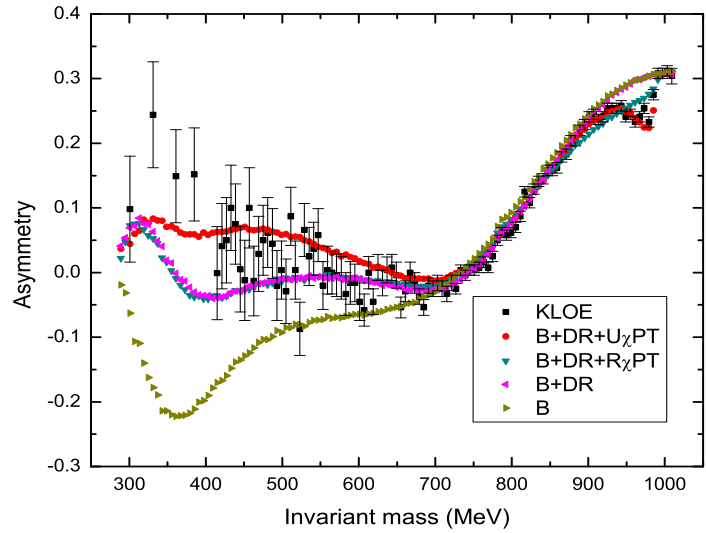}
\end{center}
\caption{Comparison of the forward backward asymmetry adding to the tree level 
bremsstrahlung plus double resonance ($B+DR$) the loop level contributions from 
resonance chiral perturbation theory ($B+DR+R\chi PT$) or unitarized chiral 
perturbation theory ($B+DR+U\chi PT$) with KLOE data \cite{KLOE,tesis}.}%
\label{UCHPT2}%
\end{figure}

\begin{table}[ptbh]%
\begin{tabular}
[c]{||c||c||c||}\hline\hline
Parameter & Value & Reference\\\hline\hline
$m_{f}$ (MeV) & $980$ & \cite{PDG}\\\hline\hline
$\Gamma_{f}$ (MeV) & $70$ & \cite{PDG} (Averaged)\\\hline\hline
$g_{f\pi^{+}\pi^{-}}$ (GeV) & $1.9$ & \cite{gallegos1}\\\hline\hline
$g_{\sigma\pi^{+}\pi^{-}}$ (GeV) & $2.4$ & \cite{achasov03}\\\hline\hline
$g_{\phi}$ & $4.42$ & \cite{PDG,LN}\\\hline\hline
$g_{fK^{+}K^{-}}$ (GeV) & $2.79$ & \cite{gallegos1}\\\hline\hline
$g_{\sigma K^{+}K^{-}}$ (GeV) & $0.55$ & \cite{achasov03}\\\hline\hline
$f_{\phi}$ & $13.3$ & \cite{PDG,LN}\\\hline\hline
$g_{\rho}$ & $5.99$ & \cite{PDG,LN}\\\hline\hline
$f_{\rho}$ & $4.96$ & \cite{PDG,LN}\\\hline\hline
\end{tabular}
\caption{Constants appearing in the $KLM$ and the numerical values used
in this work.}%
\label{tabla}%
\end{table}

\noindent Besides the intrinsic parameters of the scalar mesons, the $LSM$
involves the scalar mixing angle. For the $f_{0}(980)$ we take $m_{f}=980$
MeV, $\Gamma_{f}=70$ MeV, while for the sigma meson we use the values reported
in \cite{gallegos2} $m_{\sigma}=528$ MeV and $\Gamma_{\sigma}=414$ MeV and for
the scalar mixing angle we consider three values: $\phi_{S}=-2^{\circ}$
(LSM2), $-4^{\circ}$ (LSM4) and $-6^{\circ}$ (LSM6). 

\begin{figure}[ptbh]
\begin{center}
\includegraphics[height=3.3in,width=4.7in]{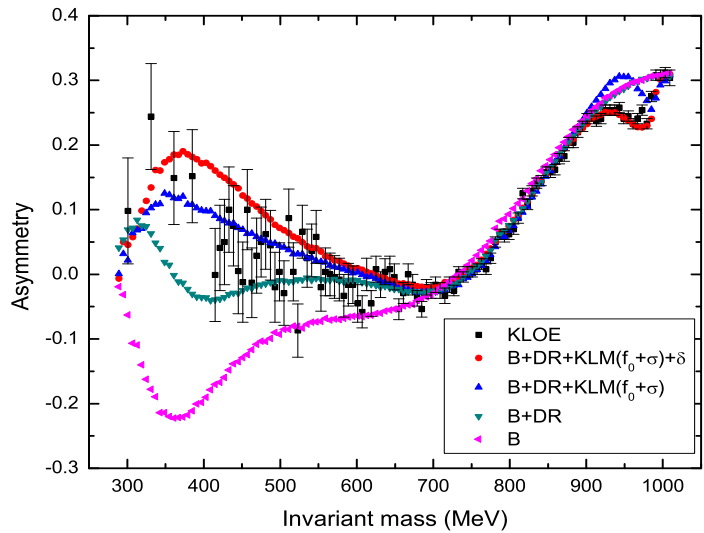}
\end{center}
\caption{Comparison of KLOE data with the results obtained adding the kaon loop 
model ($B+DR+KLM(f_{0}+\sigma)$) to the tree level bremsstrahlung plus double 
resonance exchange ($B+DR$). Points labeled $B+DR+KLM(f_{0}+\sigma)+\delta$ 
include an energy dependent phase in the kaon loop contributions proposed in 
\cite{Achasov01,Achasov02}.}%
\label{KL}%
\end{figure}

\begin{figure}[ptbh]
\begin{center}
\includegraphics[height=3.3in,width=4.7in]{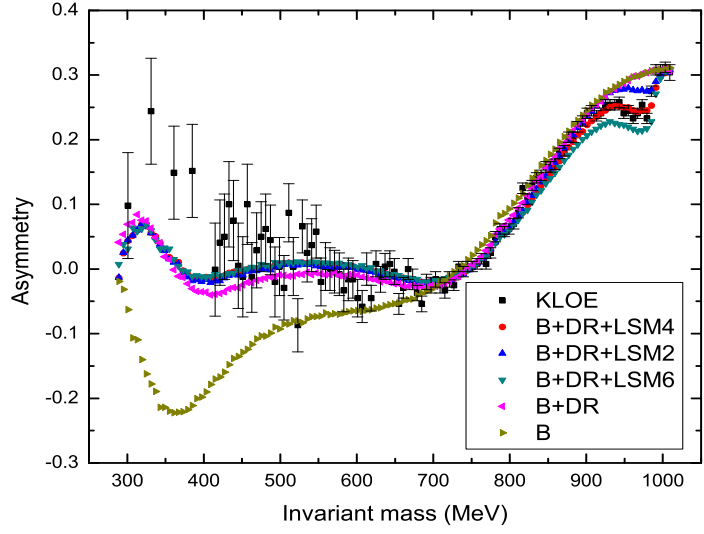}
\end{center}
\caption{Comparison of KLOE data with results for the forward backward
asymmetry as predicted by the $LSM$ using the scalar mixing angles $\phi
_{S}=-2^{\circ}$ (LSM2), $-4^{\circ}$ (LSM4) and $-6^{\circ}$ (LSM6).}%
\label{LSM01}%
\end{figure}

\begin{figure}[ptb]
\begin{center}
\includegraphics[height=3.3in,width=4.7in]{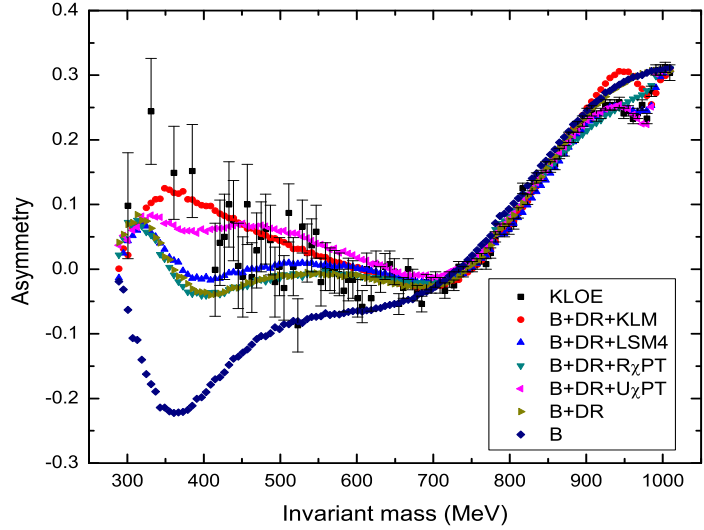}
\end{center}
\caption{Comparison of KLOE data with predictions of the four models.}%
\label{todos}%
\end{figure}

\begin{figure}[ptb]
\begin{center}
\includegraphics[height=3.3in,width=4.7in]{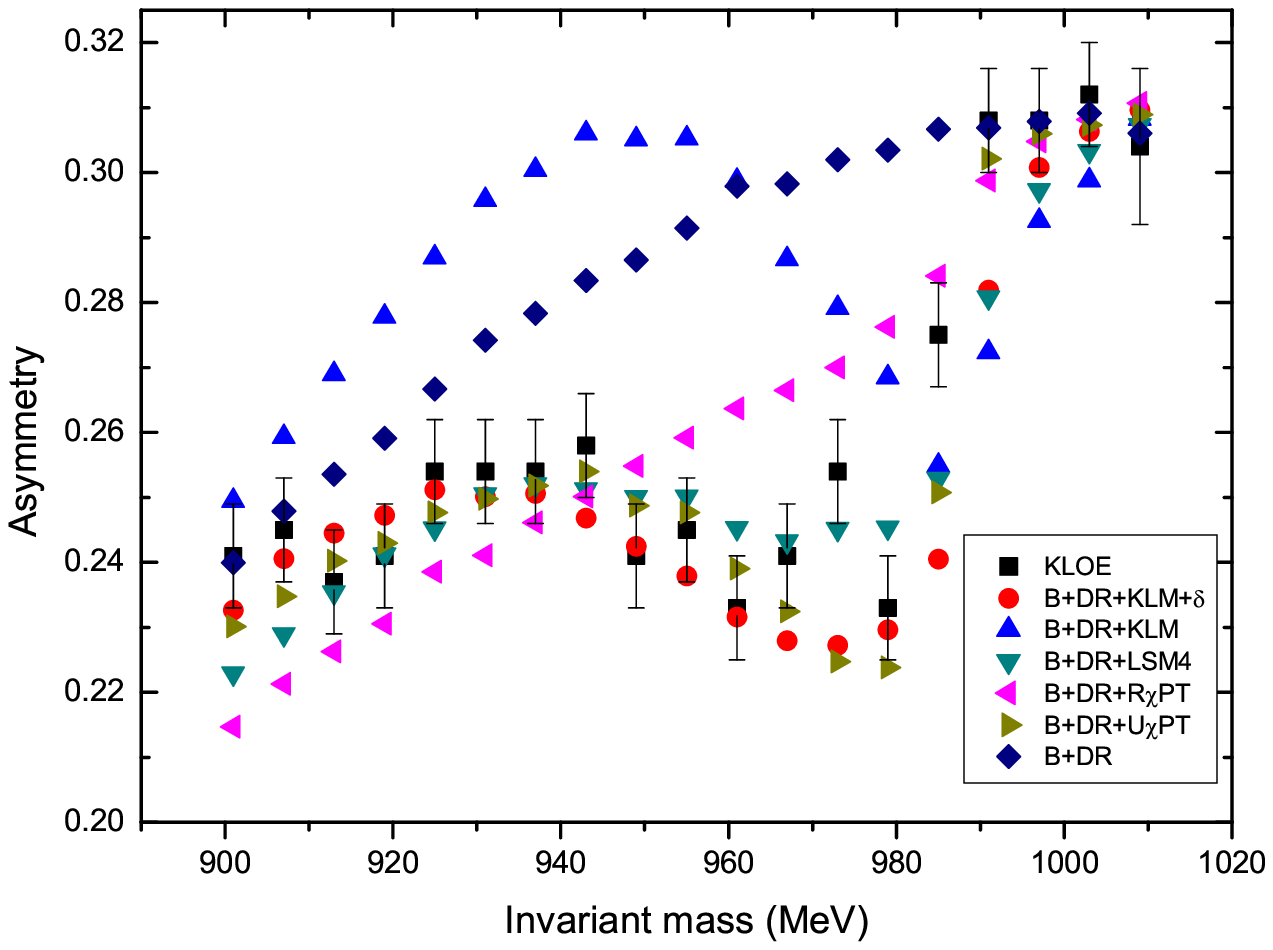}
\end{center}
\caption{Comparison of KLOE data with predictions of the four models in the
900-1020 MeV region.}%
\label{todos2}%
\end{figure}

\noindent Our results are shown in Figs. (\ref{UCHPT},\ref{UCHPT2},\ref{KL}%
,\ref{LSM01},\ref{todos},\ref{todos2}), where aiming to understand the strength 
of different of the contributions we report partial results. Below we highlight 
the main findings :

\begin{itemize}
\item There are tree level and loop contributions to the $e^{+}e^{-}%
\rightarrow\pi^{+}\pi^{-}\gamma$ process. The tree level contributions we
consider are the bremsstrahlung ($B$) and the double resonance exchange ($DR$)
shown in Figs. (\ref{Bremms},\ref{double}). The former is well known to be
dominant while the latter is expected to be small due to the $\omega-\phi$ mixing.
However this contribution is enhanced at low dipion invariant mass because of 
$\rho$ resonant effects arising in diagrams of Fig. (\ref{double}). These 
resonant effects occur when the energy of the firstly emitted pion is close 
to $210$ $MeV$, which is kinematically allowed.

\item Figure (\ref{UCHPT}) shows the bremsstrahlung ($B$), bremsstrahlung plus
double resonance exchange ($B+DR$) and bremsstrahlung plus loop level
contributions for the different models considered here ($B+U\chi PT,\,B+R\chi
PT,\,B+KLM,\,B+LSM$). bremsstrahlung alone is close to data in the 700-900 MeV
region and the remaining contributions do not modify this picture, meaning 
that they yield negligible contributions in this energy region. However, 
bremsstrahlung alone does not describe data in the sigma and $f_{0}(980)$ regions. 
Data is not reproduced at low dipion invariant mass, even if models involving 
loops are added to the bremsstrahlung contributions. We can see that double 
resonance contributions turn out to be very important at low dipion invariant mass. 
Contributions of $B+DR$ are close to data in this energy region in spite of the fact 
that $DR$ is formally sub-leading due to the $\omega - \phi$ mixing. This is due to 
the $\rho$ resonant effect mentioned above. In the $f_{0}$ region, data is well 
described by $B+U\chi PT$ and $B+LSM4$, which contain the $f_{0}$ pole but not 
by $B,\ B+R\chi PT$ or $B+DR$ where the $f_{0}$ pole is absent. Special mention 
deserves the $B+KLM$ contributions which in spite of including the $f_{0}$ pole 
it does not describe data in this region. Below we further discuss this point.

\item Figure (\ref{UCHPT2}) shows the results obtained from the $U\chi PT$ or
the $R\chi PT$ models plus the complete tree level contributions ($B+DR$).
Comparison of Figs. (\ref{UCHPT},\ref{UCHPT2}) shows that a constructive
interference between $B+DR$ and $U\chi PT$ in the sigma region takes place,
and this improves the agreement with data. At high dipion invariant mass the
$f_{0}(980)$ pole contained in the $U\chi PT$ amplitude yields the
appropriate contributions to achieve agreement with the data. Our results 
for $B+DR+U\chi PT$ agree with results recently obtained in \cite{Roca:2009zy}. 

\item Results for $B+DR+KLM$ are shown in figure (\ref{KL}). Predictions at 
low dipion invariant mass are close to data but, as mentioned above, the asymmetry 
in the $f_{0}$ region is not well described. Following \cite{Achasov01,Achasov02} 
we studied the effect of an energy dependent phase in the kaon loop amplitude. 
The authors of \cite{Achasov01,Achasov02} attribute this phase to the elastic 
background contributions and they showed this phase to be relevant for the 
interference between ISR and FSR amplitudes. In the case of charged pions 
in the final state it was extracted from data as $e^{i\delta_{B}}$ where 
$\delta_{B}=b\sqrt{q^{2}-4m_{\pi}^{2}}$ with $b=75^{\circ}/GeV$ 
\cite{kuhn02,Achasov01,Achasov02}. By including this energy dependent phase 
we obtain good agreement with data in the $f_{0}$ region and data at very low 
energy is also improved. 

\item Results for $B+DR+LSM$ are shown in figure (\ref{LSM01}). The predictions 
are not sensitive to the scalar mixing angle in the sigma region. In contrast, 
at high dipion invariant mass, the asymmetry is sensitive to this angle and the 
best agreement with data is obtained for $\phi_{S}=-4^{\circ}$. This value is 
close to the one reported in \cite{Bramon}.

\item Figure (\ref{todos}) contains a summary of the best predictions of the
models considered in this work. The detail in the 900-1020 MeV region is shown
in figure (\ref{todos2}).
\end{itemize}

\noindent Finally, we would like to remark that our aim is to explore the 
compatibility of the considered models with existing data. We find that the 
models including contributions at the loop level and containing the scalar 
poles ($U\chi PT$, $LSM$ and $KLM$) are able to reproduce the data. A detailed 
analysis of the uncertainties in the asymmetry is beyond the scope of this 
work but it will become compulsory when more precise data at low energies 
be available so that discrimination between these three models be possible. 

\section{Summary and Conclusions}

\noindent We consider the theoretical description of the $e^{+}e^{-}%
\rightarrow\pi^{+}\pi^{-}\gamma$ process. We include tree level contributions 
(bremsstrahlung and double resonance exchange) as well as loop contributions 
which are described in terms of four alternative models: Unitary Chiral 
Perturbation Theory, Resonance Chiral Perturbation Theory, Linear Sigma 
Model and Kaon Loop Model. We perform a detailed comparison of the model 
predictions and the KLOE data for the forward-backward charge asymmetry. 
Our main conclusions are listed below: 

\bigskip

\begin{itemize}
\item Bremsstrahlung accounts for data in the 700-900 MeV region.
Adding double resonance contributions brings predictions close to data
at low dipion invariant mass. \bigskip

\item The $R\chi PT$ contributions at loop level are small in the whole
kinematic region. \bigskip

\item $U\chi PT$ and $LSM$ yield an appropriate description of data when 
added to the full tree level contribution (bremsstrahlung plus double 
resonance exchange). \bigskip

\item Adding $KLM$ to the full tree level contributions yields results that 
are close to data except in the $f_{0}$ region. Agreement with data in the 
whole energy range is achieved by including an energy dependent phase proposed 
in \cite{Achasov01,Achasov02} in the $KLM$ amplitude.

\item The Linear Sigma Model predictions for the asymmetry, at high dipion
invariant mass, is highly sensitive to the scalar mixing angle, the value $\phi
_{S}=-4^{\circ}$ is favored by data. \bigskip

\item In spite of being suppressed by the $\omega-\phi$ mixing, the double 
resonance exchange shown in Fig. (\ref{double}), turns out to be crucial in 
order to describe the KLOE data in the low dipion
invariant mass. This contribution is enhanced due to resonant $\rho$ exchange.
\end{itemize}

\subsection*{Acknowledgments}

\noindent{Work supported by CONACyT under projects 50471-F and J49178-F.
Partial support from DINPO-UG is also acknowledged. We thank J. A. Oller, E.
Oset, L. Roca and C. Bini for useful suggestions.}


\begin{thebibliography}{99}                                                                                               %


\bibitem {KLOEf0}A. Aloisio \textit{et al.}, [KLOE Collaboration], Phys. Lett.
B \textbf{537}, 21 (2002) arXiv:hep-ex/0204013.

\bibitem {KLOEa0}A. Aloisio \textit{et al.}, [KLOE Collaboration], Phys. Lett.
B \textbf{536}, 209 (2002) arXiv:hep-ex/0204012.

\bibitem {KLOE}F. Ambrosino \textit{et al}. [KLOE Collaboration], Phys. Lett.
B \textbf{634}, 148 (2006) arXiv:hep-ex/0511031.

\bibitem {kuhn01}S. Binner, J. H. K\"{u}hn and K. Melnikov, Phys. Lett. B
\textbf{459}, 279 (1999) arXiv:hep-ph/9902399.

\bibitem {PSV}G. Pancheri, O. Shekhovtsova and G. Venanzoni, Phys. Lett. B
\textbf{642}, 342 (2006) arXiv:hep-ph/0605244.

\bibitem {pire}Z.~Lu and I.~Schmidt,
Phys.\ Rev.\ D \textbf{73}, 094021 (2006) [Erratum-ibid.\ D \textbf{75},
099902 (2007)] [arXiv:hep-ph/0603151];
M.~Diehl, T.~Gousset, B.~Pire and O.~Teryaev,
Phys.\ Rev.\ Lett.\ \textbf{81}, 1782 (1998) [arXiv:hep-ph/9805380];
M.~Diehl, T.~Gousset and B.~Pire,
Phys.\ Rev.\ D \textbf{62}, 073014 (2000) [arXiv:hep-ph/0003233].


\bibitem {kuhn02}H. Czyz, A. Grzelinska and H. K\"{u}hn, Phys. Lett. B
\textbf{611}, 116 (2005) arXiv:hep-ph/0412239.

\bibitem {Giulia}
S.~Dubinsky, A.~Korchin, N.~Merenkov, G.~Pancheri and O.~Shekhovtsova,
Eur.\ Phys.\ J.\ C \textbf{40}, 41 (2005) arXiv:hep-ph/0411113.


\bibitem {kuhn03}G. Rodrigo, H. Czyz, J. H. K\"{u}hn and M. Szopa, Eur. Phys.
J. C \textbf{24}, 71 (2002) arXiv:hep-ph/0112184; H. Czyz, A, Grzelinska, J.
H. K\"{u}hn and G. Rodrigo, Eur. Phys. J. C \textbf{27}, 563 (2003) arXiv:hep-ph/0212225.


\bibitem {Giulia2}G. Pancheri, O. Shekhovtsova and G. Venanzoni, J. Exp.
Theor. Phys. \textbf{106}, 470 (2008) arXiv:0706.3027 [hep-ph].


\bibitem {EGPR}
G.~Ecker, J.~Gasser, A.~Pich and E.~de Rafael,
Nucl.\ Phys.\ B \textbf{321}, 311 (1989).

\bibitem {OO}
J.~A.~Oller and E.~Oset,
Nucl.\ Phys.\ A \textbf{620}, 438 (1997) [Erratum-ibid.\ A \textbf{652}, 407
(1999)];
J.~A.~Oller, E.~Oset and J.~R.~Pelaez,
Phys.\ Rev.\ Lett.\ \textbf{80}, 3452 (1998) [arXiv:hep-ph/9803242];
J.~A.~Oller, E.~Oset and J.~R.~Pelaez,
Phys.\ Rev.\ D \textbf{59}, 074001 (1999) [Erratum-ibid.\ D \textbf{60},
099906 (1999)] [arXiv:hep-ph/9804209];
J.~A.~Oller and E.~Oset,
Phys.\ Rev.\ D \textbf{60}, 074023 (1999) [arXiv:hep-ph/9809337].


\bibitem {Levy}M. L\'{e}vy, Nuovo Cim. LIIA 23 (1967); S. Gasiorowicz, D. A.
Geffen, Rev. Mod. Phys. \textbf{41}, 531 (1969); J. Schechter, Y. Ueda, Phys.
Rev. D \textbf{3}, 2874 (1971);

\bibitem {Simon}
M.~Napsuciale,
arXiv:hep-ph/9803396;
M.~Napsuciale and S.~Rodriguez,
Int.\ J.\ Mod.\ Phys.\ A \textbf{16}, 3011 (2001) [arXiv:hep-ph/0204149].


\bibitem {LN}
J.~L.~Lucio Martinez and J.~Pestieau,
Phys.\ Rev.\ D \textbf{42}, 3253 (1990)
and J.~L.~Lucio Martinez and M.~Napsuciale,
Phys.\ Lett.\ B \textbf{331}, 418 (1994).


\bibitem {Isidori}G. Isidori, L. Maiani, M. Nicolaci and S. Pacetti, JHEP
0605:049 (2006) arXiv:hep-ph/0603241.

\bibitem {Roca:2009zy}L.~Roca and E.~Oset,
Phys.\ Rev.\ D \textbf{81}, 014010 (2010) [arXiv:0911.0994 [hep-ph]].


\bibitem {NSOV}
M.~Napsuciale, E.~Oset, K.~Sasaki and C.~A.~Vaquera-Araujo,
Phys.\ Rev.\ D \textbf{76}, 074012 (2007) [arXiv:0706.2972 [hep-ph]].


\bibitem {GVFV}
G.~Ecker, J.~Gasser, H.~Leutwyler, A.~Pich and E.~de Rafael,
Phys.\ Lett.\ B \textbf{223}, 425 (1989).

\bibitem {Bramon}A. Bramon, R. Escribano, J. L. Lucio M., M. Napsuciale, G.
Pancheri, Eur. Phys. J. C \textbf{26}, 253 (2002) arXiv:hep-ph/0204339.

\bibitem {tesis}P. Beltrame, \textit{Ph.D. Thesis} (2009), http://digbib.ubka.uni-karlsruhe.de/documents/711883.

\bibitem {PDG}C. Amsler \textit{et al.}, Phys. Lett. B \textbf{667}, 1 (2008).

\bibitem {gallegos1}R. Escribano, A. Gallegos, J. L. Lucio M, G. Moreno and J.
Pestieau, Eur. Phys. J. C \textbf{28}, 107 (2003) arXiv:hep-ph/0204338.

\bibitem {achasov03}N. N. Achasov and A. V. Kiselev, Phys. Rev. D \textbf{73}, 054029 (2006) [Erratum-ibid. D \textbf{74},059902 (2006)] [arXiv:hep-ph/0512047].

\bibitem {gallegos2}A. Gallegos, J. L. Lucio M and J. Pestieau, Phys. Rev. D
\textbf{69}, 074033 (2004) arXiv:hep-ph/0311133.

\bibitem {Achasov01}N. N. Achasov and V. V. Gubin, Phys. Rev. D \textbf{56},
4084 (1997) arXiv:hep-ph/9703367; N. N. Achasov and V. V. Gubin, Phys. Rev. D
\textbf{57}, 1987 (1998) arXiv:hep-ph/9706363.

\bibitem {Achasov02}N. N. Achasov and V. V. Gubin, Phys. Rev. D \textbf{63},
094007 (2001) arXiv:hep-ph/0101024.
\end{thebibliography}
\end{document}